\documentclass[twocolumn,showpacs,preprintnumbers,amsmath,amssymb]{revtex4}

\usepackage{epsfig}% Include encapsulated postscript figure files
\usepackage{graphicx}% Include figure files 
\usepackage{dcolumn} %Align table columns on decimal point
\def\Rthresh{R_{\mbox{\tiny\rm th}}} \def\mf{\Omega_0}
\def\sdf{\Delta\Omega}
\begin{document}

\title{Signal detection by means of phase coherence induced through
  phase resetting} \date{\today} \author{Bj\"orn Naundorf} \author{Jan
  A. Freund} \email{freund@physik.hu-berlin.de}
% \homepage{http://www.Second.institution.edu/~Charlie.Author}
\affiliation{ Institut f\"ur Physik, Humboldt-Universit\"at zu Berlin,
  Invalidenstr.~110, D-10115 Berlin, Germany }
% \altaffiliation[Also at ]{Physics Department, XYZ University.}
% Lines break automatically or can be forced with \\

\date{\today}% It is always \today, today, % but any date may be
             % explicitly specified

%%%%%%%%%%%%%%%%%%%%%%%%%%%%%%%%%%%%%%%%%%%%%%%%%%%%%%%%%%%%%%%%%%%%%%%
\begin{abstract}
  Detection and location of moving prey utilizing electrosense or
  mechanosense is a strategy commonly followed by animals which cannot
  rely on visual sense or hearing. In this article we consider the
  possibility to detect the source of a localized stimulus that
  travels along a chain of detectors at constant speed. The detectors
  are autonomous oscillators whose frequencies have a given natural
  spread. The detection mechanism is based on phase coherence which is
  built up by phase resetting induced by the passing stimulus.
\end{abstract}
%%%%%%%%%%%%%%%%%%%%%%%%%%%%%%%%%%%%%%%%%%%%%%%%%%%%%%%%%%%%%%%%%%%%%%%
%\pacs{Valid PACS appear here}% PACS, the Physics and Astronomy
                             % Classification Scheme.
%87. Biological and medical physics
%87.10.+e General theory and mathematical aspects
%87.19.Bb Sensory perceptions
%\keywords{Suggested keywords}%Use showkeys class option if keyword
                              %display desired
%%%%%%%%%%%%%%%%%%%%%%%%%%%%%%%%%%%%%%%%%%%%%%%%%%%%%%%%%%%%%%%%%%%%%%%
\maketitle
%%%%%%%%%%%%%%%%%%%%%%%%%%%%%%%%%%%%%%%%%%%%%%%%%%%%%%%%%%%%%%%%%%%%%%%
The ability to detect, locate, and capture prey is vital for survival.
Many animals accomplish these tasks using visual or acoustic
information. However, species that have developed in an environment
where these senses are obscured, have to rely on alternative
mechanisms. For example, the paddlefish ({\em Polyodon spathula}),
found in the river basins of the Midwestern United States and in the
Yangtze River in China, makes use of a passive electrosensory system
\cite{pfish}. Another example is the weakly electric fish that
combines active and passive electrosense with a mechanosensory lateral
line system \cite{elfish}. In these animals, receptors transform
stimuli into electric signals which excite the terminals of primary
afferent neurons. These afferents are well known to exhibit periodic
spike patterns \cite{spikepat}.

In the last decades a lot of research has been devoted to the details
of information processing on the neural level, i.e., the dynamics of
single neurons or neural networks. However, at the behavioral level
still many open problems exist. Since the performance and the analysis
of experiments usually involve an enormous effort, efficient and
tractable models are indispensable, both for planning and
interpretation.

Here we present an idealized, however, analytically tractable model,
proposing a mechanism for the detection of a localized stimulus. This
stimulus is passing an array of receptors, which we model as phase
oscillators. To measure the degree of coherence between the
oscillators we choose the well known synchronization index
\cite{synchroindex}. First we examine the influence of a random
initial distribution of the oscillator phases on the synchronization
index and introduce a threshold value to distinguish a stimulus from a
``false alarm''. Then we investigate the influence of our model
parameters for the detection of a moving stimulus.

We consider a linear chain of $N$ uncoupled phase rotors which are
characterized by the set of variables $\psi_1,\ldots,\psi_N$. The
rotors are aligned at equal distance along an axis of length $L$,
i.e.~the position of rotor $k$ is $x_k=(N-k)/(N-1)L$,
$k=1,\ldots,N$. Each rotor has its own natural frequency $\Omega_k$
which, in the absence of a stimulus, determines the simple linear
growth of the phase, i.e.~$\psi_k(t)=\psi_k(0)+\Omega_k t$. We assume
the frequencies to be independently and identically distributed
according to a Gaussian with mean $\mf$ and standard deviation
$\sdf=\eta\mf$.

An appropriate quantity to measure the degree of phase coherence among
these rotors is the complex variable
\begin{equation}
Z(t)=\frac{1}{N}\sum_{k=1}^N \exp\left[i\psi_k(t)\right] =:
R(t)\exp\left[i\Phi(t)\right].
\end{equation}
This global order parameter contains both the information about the
instantaneous collective phase $\Phi(t)$ and the instantaneous degree
of phase coherence measured by the modulus $R(t)$ at time $t$. Its
square can be expressed in several ways:
\begin{eqnarray}
\label{rsq1}
R^2(t) &=& \frac{1}{N^2} \left\{ \left[\sum_{k=1}^N
\cos(\psi_k(t))\right]^2 + \left[\sum_{k=1}^N
\sin(\psi_k(t))\right]^2\, \right\}\\[.4cm]
\label{rsq2}
&=& \frac{1}{N^2} \left[ \sum_{k,l=1}^N \cos(\psi_k(t)-\psi_l(t))
\right]\\[.4cm]
\label{rsq3}
&=& \frac{1}{N} + \frac{2}{N^2} \sum_{k=1}^N \sum_{l>k}^N
\cos(\psi_k(t)-\psi_l(t))
\end{eqnarray}
This quantity is termed synchronization index since it is widely used
in the description of synchronization processes \cite{synchroindex}.
From Eqs.~(\ref{rsq1}) and (\ref{rsq2}) it is obvious that $0\le
R(t)\le 1$ with $R(t)=1$ indicating perfect coherence.

\begin{figure}
  \centerline{\psfig{figure=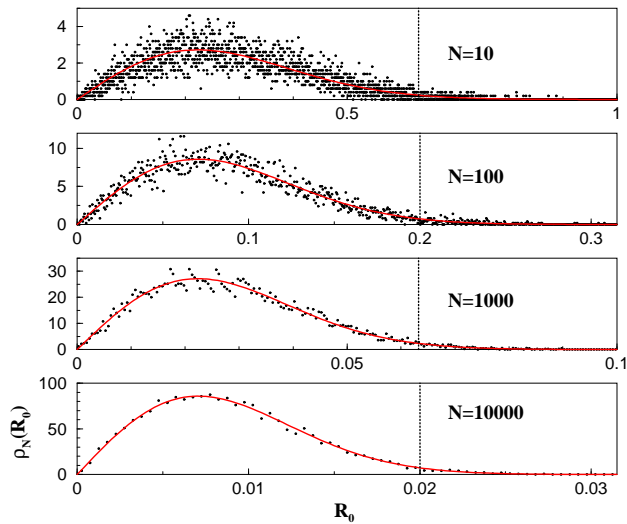,width=8.3cm}} \caption{The
  density of the random variable $R_0$ for different
  $N=10,100,1000,10000$ (top down). The dots indicate the result of
  numerical simulations whereas the full line identifies the Rayleigh
  distributions (\ref{eq:Maxi}). Vertical lines mark the threshold set
  by the value $2/\sqrt{N}$, which corresponds to a $2\%$ level of
  false alarm.}  \label{fig:init_histos}
\end{figure}

We initialize the array by randomly selecting a phase for each of the
rotors according to the uniform distribution on $[0,2\pi)$.  Thus, the
quantity $R_0=R(t=0)$ is a random variable. Its density contains
important information because even in the absence of any signal the
array of rotors will generate nonvanishing values of $R_0$. These
have to be discriminated from values of $R(t)$ which significantly
indicate coherence induced by the passing stimulus.
Figure \ref{fig:init_histos} shows numerically estimated densities, where
$N=10,100,1000,10000$ equidistributed phases were used to compute a
single realization of the random variable $R_0$.
An analytic expression for these distributions can be derived by
applying the central limit theorem (Lindenberg-L\'evy theorem) to the
following pair of random variables
\begin{equation}
  \label{eq:ranvars} X_N={1\over N}\sum_{k=1}^N \cos[\psi_k(0)]\,,
  \quad Y_N={1\over N}\sum_{k=1}^N \sin[\psi_k(0)] \end{equation}
  which yields for large $N$ the limiting density \cite{Fisz}
\begin{equation}
  \label{eq:pxy} \rho(X_N=x,Y_N=y) \simeq {N\over \pi}
  \exp\left[-N(x^2+y^2)\right]\,.
\end{equation}
Changing to polar coordinates $(R_0,\Phi_0)$ and integrating over the
angle $\Phi_0$ immediately leads to the Rayleigh distribution
(Fig.~\ref{fig:init_histos})
\begin{equation}
  \label{eq:Maxi} \rho_N(R_0)=2NR_0\exp(-NR_0^2)\,.
\end{equation}
Mean and variance of this distribution read
\begin{equation}
  \label{eq:mf_sdf} \left<R_0\right>={\sqrt{\pi}\over
  2}{1\over\sqrt{N}}\,,\qquad \left<\Delta
  R_0^2\right>=\left[1-{\pi\over 4}\right]{1\over N}\,,
\end{equation}
The integral
\begin{equation}
  \label{eq:quantile} \alpha(\Rthresh)=
  \int\limits_{\Rthresh}^{\infty} 2NR_0\exp(-NR_0^2)\;dR_0 =
  \exp(-N\Rthresh^2)
\end{equation}
can be used to define a threshold value $\Rthresh$ by demanding that
$\alpha(\Rthresh)$, which is the probability for false alarm, be less
than some fixed small number. For instance, $\Rthresh=2/\sqrt{N}$
corresponds to $\alpha(\Rthresh)\le 2\%$ which means values larger
than $2/\sqrt{N}$ occur by random configuration with a probability of
less than 2\%. In what follows we will use $\Rthresh=2/\sqrt{N}$ to
discriminate stimuli against the random configuration background.

A standard model describing phase resetting by an external stimulus of
strength $I$ is given by the following phase dynamics \cite{Tass01}:
\begin{equation}
  \label{eq:standmod} \dot\psi = \Omega + I\cos \psi\,.
\end{equation}
This dynamics can be illustrated as the overdamped motion in a tilted
corrugated potential landscape. If $I<\Omega$ no troughs (minima) and
barriers (maxima) exist and the phase continues cycling forward
($\Omega>0$) at varying speed. For $I>\Omega$ two fixed points emerge,
which correspond to a minimum at $\pi-\arccos(\Omega/I)$ and a maximum
at $\pi+\arccos(\Omega/I)$. For constant $I$ the phase settles in the
minimum (mod$2\pi$) regardless of the initial position, which means
the phase eventually is reset to the corresponding value
$\psi^*=\pi-\arccos(\Omega/I)$. The situation is harder to analyze
with a time varying stimulus $I(t)$; the net effect will depend on
many details of the stimulus, e.g., the time scale of variation, the
height of the signal peak, etc.

\begin{figure}
  \centerline{\psfig{figure=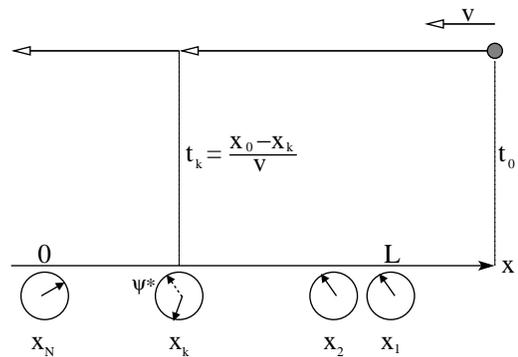,width=7.cm}} \caption{Sketch
  of the setup. The stimulus moves at constant speed $v$ relative to
  the oscillator chain. Each time it passes an oscillator the phase is
  reset to $\psi^*$.}  \label{fig:sketch}
\end{figure}

Our detection setup would require to consider $N$ such phase equations
each with its own time varying stimulus $I_k(x_0-x_k-vt)$, where $x_0$
and $v$ are the initial position (at time $t=0$) and the constant
velocity of the traveling stimulus, respectively. Irrespective of the
details, the equation of motion will be too complicated to be solved
analytically in closed form. If, however, the peak value of the
stimulus is sufficiently high and the duration is short, we can
simplify the resetting mechanism: The passing stimulus resets the
phase $\psi_k(t)$ to some global value $\psi^*$ the very moment it is
{\em at} position $x_k$, i.e., the reset is instantaneous.  After this
reset the phase again increases linearly with its natural frequency
$\Omega_k$. The situation is sketched in Fig.~\ref{fig:sketch}.
The history of phase $\psi_k$ can thus be written as
\begin{equation}
  \label{eq:hist} \psi_k(t)=\left\{ {\psi_k(0)+\Omega_k t\qquad t<t_k
  \atop \psi^*+\Omega_k (t-t_k) \quad t\ge t_k }\right.
\end{equation}
(for all $k=1,\ldots,N)$, where $t_k$ is the time when the stimulus
passes the oscillator $k$. Substituting this into Eq.~(\ref{rsq3}) we
find the following value of the synchronization index:
\begin{equation}
  \label{eq:r2a} R^2(t) = {1\over N}+{2\over
  N^2}\left\{S_{kk}(t) + S_{kN}(t) + S_{NN}(t)\right\}
\end{equation}
in the time interval $t_k\le t<t_{k+1}$, where we denote
\begin{eqnarray}
  \label{eq:Swhat} S_{kk}(t) &=& \sum_{i=1}^k\sum_{j>i}^k
  \cos\left[\Omega_i(t-t_i)-\Omega_j(t-t_j)\right], \\[2mm] S_{kN}(t)
  &=& \sum_{i=1}^k\sum_{j>k}^N
  \cos\left[\psi^*+\Omega_i(t-t_i)-\psi_j(0)-\Omega_j t\right], \\[2mm]
  S_{NN}(t) &=& \sum_{i=k+1}^N\sum_{j>i}^N
  \cos\left[\psi_i(0)+\Omega_i t -\psi_j(0)-\Omega_j t\right]\,.
\end{eqnarray}
These expressions depend on the initial phases $\psi_i(0)$ and the
natural frequencies $\Omega_i$. We consider both quantities to be
random parameters of the model. To characterize the net effect of
observing many realizations, i.e., to evaluate the mean performance of
many individuals, we average the synchronization index over both the
initial phases (equidistributed) and the natural frequencies
(Gaussian). The first average over the phases yields (for
$t_k\leq t\leq t_{k+1}$)
\begin{equation}
  \label{eq:phaseaverage} \left<R^2(t)\right> = {1\over N}+{2\over
  N^2} \sum_{i=1}^k\sum_{j>i}^k
  \cos\left[\Omega_i(t-t_i)-\Omega_j(t-t_j)\right]\,.
\end{equation}
Note that the value of $\psi^*$ is irrelevant for this expression.
The second average over the natural frequencies results in
\begin{eqnarray}
  \label{eq:freqaverage} \left<\left<R^2(t)\right>\right> &=& {1\over
  N}+{2\over N^2} \sum_{i=1}^k\sum_{j>i}^k
  \cos\left[\Omega_0(t_j-t_i)\right]\nonumber\\[2mm] && \times\;
  \exp\left\{ -{\eta^2\Omega_0^2\over
  2}\left[(t-t_i)^2+(t-t_j)^2\right] \right\}.
\end{eqnarray}
In the following we relate time to the position of the stimulus
$x(t)$,
\begin{equation}
  t={x_0-x(t)\over v}.
\end{equation} 
We can then derive an expression that
reflects how the twice averaged global synchronization index varies as
a function of the position of the stimulus over the linear detector
chain, namely,
\begin{eqnarray}
  \label{eq:rostrum1} \left<\left<R^2(x)\right>\right>%(x_{k+1}<x\le
  &=& {1\over N}+{2\over N^2} \sum_{i=1}^k\sum_{j>i}^k
  \cos\left[{\Omega_0\over v}(x_i-x_j)\right]\nonumber\\[2mm] &&
  \times\; \exp\left\{ -{\eta^2\Omega_0^2\over
  2v^2}\left[(x-x_i)^2+(x-x_j)^2\right] \right\},
\end{eqnarray}
for $x_{k+1}<x\le x_k$.  Assuming the oscillators to be distributed
along the linear chain in an equidistant manner, i.e., $x_k=(N-k)\Delta
x$ for $k=1,\ldots,N$ with $\Delta x=L/(N-1)$, we find
\begin{eqnarray}
  \label{eq:rostrum2} \left<\left<R^2(x)\right>\right> &=& {1\over
  N}+{2\over N^2} \sum_{i=1}^{k(x)}\sum_{j>i}^{k(x)}
  \cos\left[(j-i)\kappa\right]\nonumber\\[2mm] && \times\; \exp\left\{
  -{\eta^2\kappa^2\over 2}g(x,i,j) \right\}\,,
\end{eqnarray}
where
\begin{equation}
  \label{eq:g} g(x,i,j) = \left[{x\over\Delta x} -(N-i)\right]^2 +
  \left[{x\over\Delta x} -(N-j)\right]^2
\end{equation}
and $k(x)=\min\left\{\mbox{\rm int}[N-{x\over\Delta x}],N\right\}$.
The parameter $\kappa$ turns out to be related to the ratio of the
travel time between two neighboring oscillators $\Delta T = \Delta
x/v$ and the mean rotation period $T_0=2\pi/\Omega_0$,
i.e., $\kappa=2\pi{\Delta T\over T_0}$. It is useful to write
$\kappa=2\pi m+\delta$, where $m\in N$ and $\delta\in[0,2\pi)$.

\begin{figure}
  \centerline{\psfig{figure=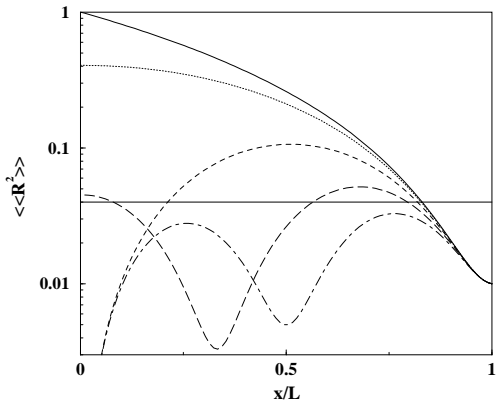,width=6.5cm}}
  \caption{Variation of the global synchronization index as a function
  of the stimulus position $x$ in the case of vanishing frequency
  spread $\eta=0$ for $N=100$. Depending on the detuning parameter
  $\delta$ constructive or destructive effects of the array or rotors
  can be observed: $\delta/\pi=0, 0.01, 0.02, 0.03, 0.04$ shown as
  solid, dotted, dashed, long-dashed, dot-dashed lines,
  respectively. The solid horizontal line marks $\Rthresh^2=4/N$}
  \label{fig:curves}
\end{figure}

\begin{figure}[htbp]
  \centerline{\psfig{figure=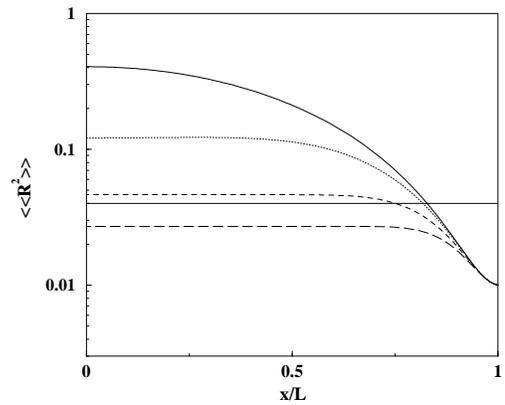,width=6.5cm}} \caption{Spatial
  variation of the global synchronization index as a function of the
  stimulus position $x$ in the case of varying frequency spread $\eta$
  for slight detuning, i.e., $\delta/\pi=0.01$, and for $m=1$ and
  $N=100$. An increasing spread erodes the detection mechanism:
  $\eta=0, 0.005, 0.01, 0.015$ shown as solid, dotted, dashed,
  long-dashed, respectively.  The solid horizontal line marks
  $\Rthresh^2=4/N$} \label{fig:spread}
\end{figure}

Equations~(\ref{eq:rostrum2}) and (\ref{eq:g}) present the central
result of our model.
\begin{figure}
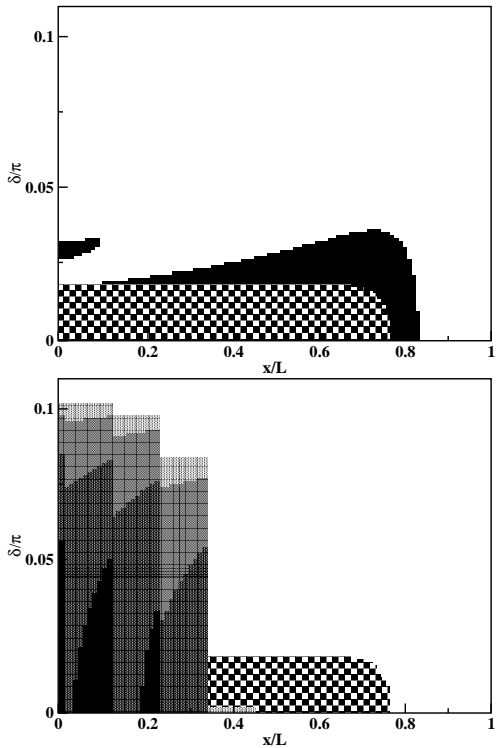

  \centerline{\psfig{figure=figure3a.eps,width=6.5cm}}
  \centerline{\psfig{figure=figure3b.eps,width=6.5cm}}
  \caption{Regions in which prey is detected in the $x$-$\delta$ plane
  defined by the demand that
  $\left<\left<R^2(x)\right>\right>>\Rthresh^2=4/N$. Top panel for
  $N=100$, $\eta=0$ (black) and $\eta=0.01$ (checkered plotted on
  top).  The bottom panel compares the region for $N=10$ and $\eta=0,
  0.01, 0.02, 0.03$ (light to dark gray plotted on top of each other)
  with $N=100$ and $\eta=0.01$ (checkered). Each time the stimulus
   passes an detector, $\left<\left<R(x)\right>\right>$ changes discontinously and decays for
   $\eta>0$.}  \label{fig:regions}
\end{figure}
The detection regions in the $x$-$\delta$ plane, i.e.~where
$\left<\left<R^2(x_k)\right>\right>$ is larger than the threshold
value $\Rthresh^2=4/N$, is shown in Fig.~\ref{fig:regions} for $N=100$
(top) and $N=10$ (bottom). It can be seen that detection works only as
long as detuning, quantified by $\delta$, and frequency spread, coded
by $\eta$, are not too large.  Moreover, we find that the detection region
shrinks in the $\delta$ direction, but enlarges in the $x$ direction
with increasing  $N$, i.e., detection already works when the stimulus has
passed only a small number of oscillators.  For small $\eta$ and small
$\delta$ we can consider the following limiting cases: First let us
deal with the case of zero frequency spread, i.e., $\eta=0$. The double
sum over cosines can be performed yielding the following expression,
\begin{equation}
  \label{eq:sol} \left<\left<R^2(x_k)\right>\right>_{\eta=0} =
  \frac{N-k+\frac{1-\cos(k\delta)}{1-\cos(\delta)}}{N^2},
\end{equation}
which we exemplify for $N=100$ in Fig.~\ref{fig:curves}.
Depending on the detuning parameter $\delta$, constructive or
destructive effects show up.
Introducing the frequency spread $\eta>0$ erodes both the constructive
and destructive effects. Note that the cycle number $m$ matters if
$\eta>0$ whereas it is irrelevant for the case $\eta=0$. In
Fig.~\ref{fig:spread} we exemplify how the detection curve for $N=100$,
 $m=1$ and a detuning value of $\delta=0.01\pi$ is pushed below the
detection threshold by an increasing frequency spread $\eta$.
These results indicate that the detection mechanism is rather
sensitive with respect to the width of the frequency distribution for
a large number $N$ of oscillators. However, we would like to point
out, that the biological relevance
is not eradicated by this finding, since evolutionary optimization
offers an explanation how the confined parameter range might have been
realized.

In conclusion, we have presented a simplified but analytically
tractable model for signal detection, which works by creating
significant coherence in a chain of phase oscillators. This coherence
is induced by a strongly localized stimulus that travels at constant
speed and resets phases instantaneously.  The ability to detect a
stimulus rapidly is balanced by the sensitivity to variations in the
oscillator frequencies or deviations from the optimal velocity. The
variations in the frequencies, however, guarantee a fast
desynchronization after the stimulus has passed.

Although our approach concentrates on seemingly crude assumptions, it
catches the main features of prey detection. Future experimental
studies have to reveal in which direction this model has to be
extended to account for given biological applications.

We thank A.~Neiman, L.~Wilkens and M.~Timme for useful
discussion. J.F.~acknowledges support by the DAAD (NFS-Project
No.~D/0104610). This work has been supported by the DFG, SFB 555.

%%%%%%%%%%%%%%%%%%%%%%%%%%%%%%%%%%%%%%%%%%%%%%%%%%%%%%%%%%%%%%%%%%%%%%%
\bibliographystyle{unsrt}

%%%%%%%%%%%%%%%%%%%%%%%%%%%%%%%%%%%%%%%%%%%%%%%%%%%%%%%%%%%%%%%%%%%%%%%
%%%%%%%%%%%%%%%%%%%%%%%%%%%%%%%%%%%%%%%%%%%%%%%%%%%%%%%%%%%%%%%%%%%%%%%
%%%%%%%%%%%%%%%%%%%%%%%%%%%%%%%%%%%%%%%%%%%%%%%%%%%%%%%%%%%%%%%%%%%%%%%

\end{document}